\documentclass{ws-procs975x65}

\begin{document}

\title{ CONSTRAINT RELAXATION}

\author{PEDRO MARRONETTI}

\address{Florida Atlantic University,\\
777 Glades Rd, 
Boca Raton, FL 33431, USA\\
\email{pmarrone@fau.edu}}

\begin{abstract}
Full relativistic simulations in three dimensions are known to
develop runaway modes that grow exponentially and are accompanied by
violations of the Hamiltonian and momentum constraints.
We present here a method that controls the violation of these constraints and
is tested with simulations of binary neutron stars in
circular orbits. We show that this technique improves the overall 
quality of the simulations.
\end{abstract}

\bodymatter

\section{Introduction}\label{intro}
For most of the past decade, the main theoretical thrust
in gravitational research has been directed toward
obtaining stable and accurate numerical models of compact-object
binary systems. One of the most difficult problems to tackle has
been the control of exponentially growing instabilities that degrade
the quality of any simulation and, eventually, terminate it.
We propose in Refs. ~\cite{Marronetti:2005bz,Marronetti:2005aq}
an evolution scheme (Constraint relaxation or CR) where the Hamiltonian and 
momentum constraints are only approximately solved at every time step, gently 
steering the evolution toward the space of solutions of the Einstein Field 
Equations without completely forcing their numerical satisfaction. This method utilizes
the conformal decomposition of the spatial metric and extrinsic curvature, 
which has traditionally been used to solve the initial value problem for binary systems.
In this decomposition, a conformal factor $\psi$ factored out of the spatial
metric and a longitudinal addition to the extrinsic curvature generated
from a vector potential $w_i$ are used to satisfy the Hamiltonian and
the three components of the momentum constraint respectively. CR
drives $\psi$ to the solution space of the Hamiltonian constraint
by means of a parabolic equation for the conformal factor.
Similarly, the momentum constraint is controlled by the use of $w_i$ to
push the simulation toward the space of solutions of the momentum
constraint. In both cases, a full relaxation of $\psi$ and $w_i$
would lead to the numerical solution of the constraints. However, the
stability of the relaxation methods relies on gently updating these
fields during the evolution. We showed in Ref. \cite{Marronetti:2005aq} 
that a full relaxation scheme becomes unstable rather quickly when used 
in combination with BSSN.

\section{Results}\label{dicu}

\begin{figure}
\begin{center}
\caption{ Evolution of the $L_2$ norm of the Hamiltonian constraint violation. }
\label{HC_BSSN_vs_HR}
    \includegraphics[width=3.6in]{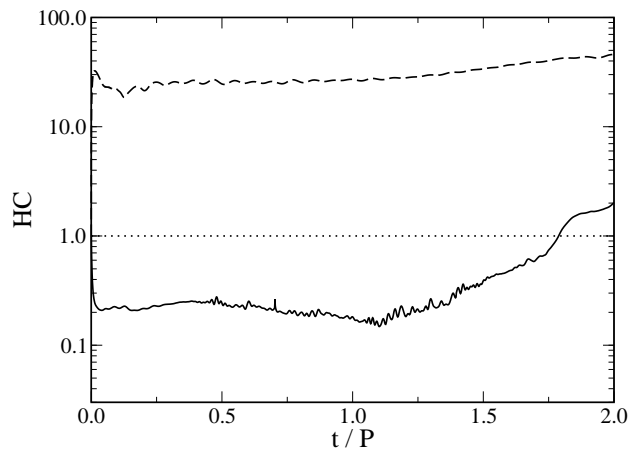}
\end{center}
\end{figure}

\begin{figure}
\begin{center}
\caption{ Evolution of the $L_2$ norm of the momentum constraint violation. }
\label{Mi_Long_Run}
    \includegraphics[width=3.6in]{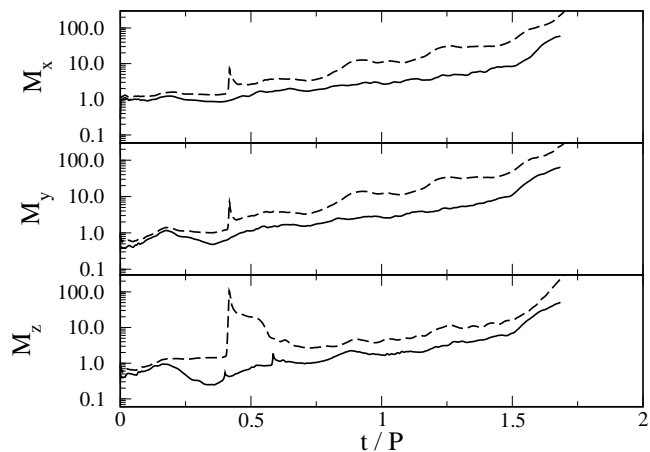}
\end{center}
\end{figure}

We tested the CR scheme with simulations
of an irrotational BNS system. The details of the initial data set
are provided in table 1 of Ref. ~\cite{Marronetti:2005bz} and the 
corresponding convergence tests are presented in Ref. ~\cite{Marronetti:2005aq}.  
The plots show curves that, for clarity, have been normalized 
to their corresponding initial values. 

The evolution of the $L_2$ norm of the Hamiltonian constraint residual
$\mathcal{HC}$ and the three components of the momentum
constraint residual $\mathcal{M}_i$ across the numerical grid are shown
in Figs. \ref{HC_BSSN_vs_HR} and \ref{Mi_Long_Run} respectively.  
CR achieves an impressive reduction of the Hamiltonian contraint violation
(Fig.\ref{HC_BSSN_vs_HR}). Note that the curves are plotted in 
logarithmic scale to highlight the more than two orders of magnitude 
difference between the CR (dashed) and BSSN (solid) results. CR 
not only suppresses the constraint 
violation modes, but also reduces the violation present in the ID 
set by a factor of about 5. The difference between the CR (solid)
and the BSSN (dashed) results in the case of the
momentum constraint (Fig. \ref{Mi_Long_Run}) is not that impressive. 
At the end of the simulation, the
momentum constraint violation was about four times smaller than in the BSSN runs.
The spikes present in the BSSN curves at $t \simeq 0.4 P$ occur on the stellar 
surface and are related to matter displacement in the grid, a side-effect of 
using a frozen shift vector. Note, however, that those spikes disappear when using 
CR.

\begin{figure}
\begin{center}
\caption{ Evolution of the total angular momentum $J$.}
\label{J_Long_Run}
    \includegraphics[width=3.6in]{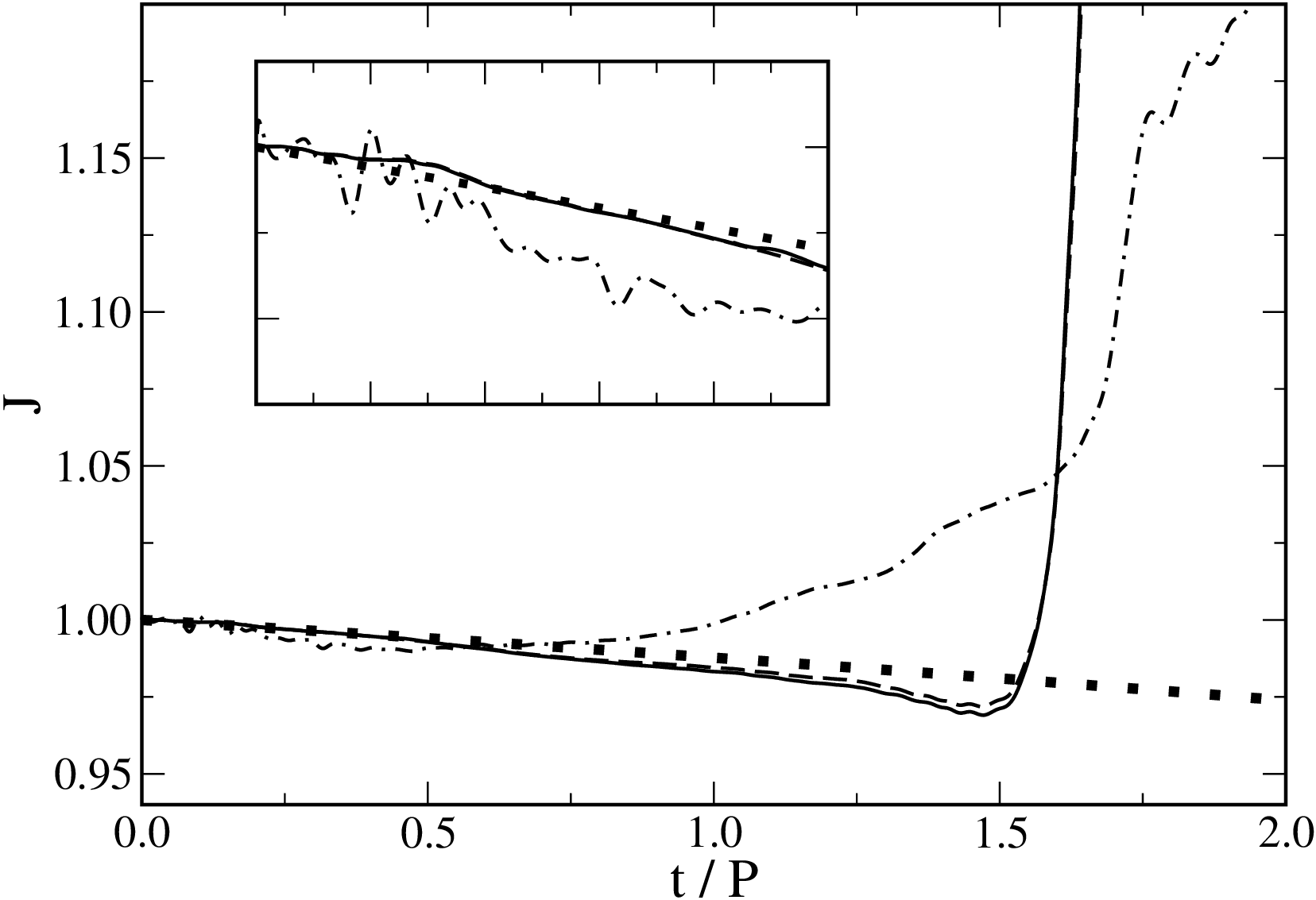}
\end{center}
\end{figure}

Figure \ref{J_Long_Run} shows the evolution of the total angular
momentum. The plot shows the CR (solid), BSSN (dashed) curves, as well as the 
PN estimation (dotted line) of the angular momentum loss for a point-mass 
binary with the same mass and angular momentum as the BNS in consideration 
(see Appendix B of Ref. ~\cite{Marronetti:2005bz}). The CR based run agrees with the 
PN prediction for about 1.5 orbital periods. The inset of figure 
\ref{J_Long_Run} zooms in on the first half of the period, showing the 
reduced level of noise in the CR curve.

\section*{Acknowledgments}
This work was partially supported by the National Science Foundation under
grant PHY-0555644 and the National Computational Science Alliance 
under grants PHY020007N. PHY050010T, and PHY050015N.

\vfill

\end{document}